\documentclass[apj]{emulateapj}
\slugcomment{Version accepted for publication in ApJL, 26 June 2006}
\usepackage{graphicx,amssymb,amsmath,times}

\begin{document}
\title{Very High Energy Gamma-ray Radiation from the Stellar-mass Black Hole Cygnus X-1}
\shorttitle{VHE gamma-ray emission from the galactic black hole Cygnus X-1}
\shortauthors{J. Albert \emph{et al.}}

%
\author{
 J.~Albert\altaffilmark{a}, 
 E.~Aliu\altaffilmark{b}, 
 H.~Anderhub\altaffilmark{c}, 
 P.~Antoranz\altaffilmark{d}, 
 A.~Armada\altaffilmark{b}, 
 C.~Baixeras\altaffilmark{e}, 
 J.~A.~Barrio\altaffilmark{d},
 H.~Bartko\altaffilmark{f}, 
 D.~Bastieri\altaffilmark{g}, 
 J.~K.~Becker\altaffilmark{h}
% ...
}
 \altaffiltext{a} {Universit\"at W\"urzburg, D-97074 W\"urzburg, Germany}
 \altaffiltext{b} {IFAE, Edifici Cn., E-08193 Bellaterra (Barcelona), Spain}
 \altaffiltext{c} {ETH Zurich, CH-8093 Switzerland}
 \altaffiltext{d} {Universidad Complutense, E-28040 Madrid, Spain}
 \altaffiltext{e} {Universitat Aut\`onoma de Barcelona, E-08193 Bellaterra, Spain}
 \altaffiltext{f} {Max-Planck-Institut f\"ur Physik, D-80805 M\"unchen, Germany}
 \altaffiltext{g} {Universit\`a di Padova and INFN, I-35131 Padova, Italy}  
 \altaffiltext{h} {Universit\"at Dortmund, D-44227 Dortmund, Germany}
% ...

\begin{abstract}
We report on the results from the observations in very high energy
band (VHE, $E_\gamma \ge 100$ GeV) of the black hole X-ray binary
(BHXB) Cygnus X-1. The observations were performed with the MAGIC
telescope, for a total of 40 hours during 26 nights, spanning the
period between June and November 2006. Searches for steady
$\gamma$-ray signals yielded no positive result and upper limits to
the integral flux ranging between 1 and 2$\%$ of the Crab nebula flux,
depending on the energy, have been established. We also analyzed each
observation night independently, obtaining evidence of $\gamma$-ray
signals at the 4.0 standard deviations ($\sigma$) significance level
(3.2$\sigma$ after trial correction) for 154 minutes effective on-time
(EOT) on September 24 between 20h58 and 23h41 UTC, coinciding with an
X-ray flare seen by {\it RXTE}, {\it Swift} and {\it INTEGRAL}. A
search for faster-varying signals within a night resulted in an excess
with a significance of 4.9$\sigma$ (4.1$\sigma$ after trial
correction) for 79 minutes EOT between 22h17 and 23h41 UTC. The
measured excess is compatible with a point-like source at the position
of Cygnus X-1, and excludes the nearby radio nebula powered by its
relativistic jet. The differential energy spectrum is well fitted by
an unbroken power-law described by $dN/(dA~dt~dE) = (2.3\pm 0.6)
\times 10^{-12} (E/1 \textrm{TeV})^{-3.2\pm 0.6}$. This is the first
experimental evidence of VHE emission from a stellar-mass black hole,
and therefore from a confirmed accreting X-ray binary.

\end{abstract}


\section{Introduction}

Cygnus X-1 is the best established candidate for a stellar
mass black-hole (BH) and one of the brightest X-ray sources in the sky
\citep{Bowyer1965}. Located at a distance of $2.2\pm 0.2$~kpc, it is composed
of a $21\pm 8$~M$_\odot$ BH turning around an O9.7~Iab companion of $40\pm
10$~M$_\odot$ \citep{Ziolkowski2005} in a circular orbit of 5.6 days
and inclination between 25$^\circ$ and 65$^\circ$
\citep{Gies1986}. The X-ray source is thought to be powered mainly by 
accretion and displays the canonical high/soft and low/hard X-ray
spectral states depending on the accretion rate
\citep{Esin1998}. The thermal soft component is produced by the
accretion disk close to the BH, whereas hard X-rays are thought to be
produced by inverse Compton scattering of soft photons by thermal
electrons in a corona or at the base of a relativistic jet.  The
results from observations in the soft $\gamma$-ray range with COMPTEL
\citep{Mcconnell2002} and {\it INTEGRAL} \citep{Cadolle2006} strongly
suggest the presence of a higher energy non-thermal component. In
addition, fast episodes of flux variation by a factor between 3 and 30
have been detected at different time scales, ranging from milliseconds
in the 3-30 keV band \citep{Gierlinski2003} to several hours in the
15-300 keV band \citep{Golenetskii2003}. Radio emission stays at a
rather stable level during the low/hard state, except for rarely
observed flares \citep{Fender2006}, and appears to be quenched below a
detectable level during the high/soft state
\citep{Brocksopp1999}. On the other hand, VLBA images have shown the
presence of a one-sided, elongated radio structure (15 mas length)
during the hard state \citep{Stirling2001}, indicating the presence of
a highly collimated (opening angle $<2^\circ$) relativistic ($v \ge
0.6c$) jet. \citet{Romero2002} have suggested that Cygnus X-1 is a
microblazar, where the jet axis is roughly aligned with the line of
sight. The interaction of the outflow from the jet with the
interstellar medium appears to produce a large-scale ($\sim 5$ pc
diameter), ring-like, radio emitting structure \citep{Gallo2005},
which implies that most of the energy from the system is released by a
radiatively inefficient relativistic jet.

Three other binary systems have been detected so far in the VHE
domain, namely PSR~B1259$-$63 \citep{Aharonian2005a}, LS~I+61~303
\citep{Albert2006a} and LS~5039 \citep{Aharonian2005b}.
In  PSR~B1259$-$63 the TeV emission is thought to be produced by the
interaction of the relativistic wind from a young non-accreting pulsar
with that of the companion star. Recent results suggest that
LS~I~+61~303 also contains a non-accreting neutron star
\citep{Dhawan2006}, while the situation is not yet clear in the case
of LS~5039. As of
now, there is no experimental evidence of VHE emission from any
galactic BHXB system.

In this letter we report on the --to our knowledge-- first results of
observations of Cygnus X-1 in the VHE regime, performed with the Major
Atmospheric Gamma Imaging Cherenkov (MAGIC) telescope. Our results
pose stringent upper limits to a steady VHE flux and include evidence
of an intense, fast flaring episode occurring in coincidence with an
X-ray flare. We briefly describe the observations and data analysis,
derive the spatial and spectral features of the observed excess, and
discuss the obtained results.

\begin{deluxetable}{rrrrrr}
\tablewidth{0pt}
\tablecaption{Cygnus X-1 observation log\tablenotemark{a}
\label{table:log}}
\tablehead{
\colhead{MJD} & \colhead{T} &\colhead{$N_\textrm{excess}$} &
\colhead{S} & \colhead{Post} & \colhead{U.L.} \\ 
\colhead{$[$days$]$}&\colhead{[min]} &\colhead{[evts]} &\colhead{[$\sigma$]}&\colhead{[$\sigma$]}&\colhead{[evts ($\%$ CU)]}
}
\startdata
 53942.051&     61.1&     3.6$\pm$  4.8&     0.8&     $<0.1$&   15.02(11.1)\\
 53964.887&    105.6&     4.8$\pm$  6.9&     0.7&     $<0.1$&   21.49(9.2)\\
 53965.895&    195.3&   $-$13.2$\pm$ 10.1&    $-$1.3&     $<0.1$&    8.74(2.0)\\
 53966.934&    124.8&     9.4$\pm$  9.5&     1.0&     $<0.1$&   33.07(11.9)\\
 53967.992&     48.5&    $-$9.0$\pm$  4.7&    $-$1.7&     $<0.1$&    1.57(1.5)\\
 53968.883&    237.5&    $-$4.4$\pm$ 11.6&    $-$0.4&     $<0.1$&   22.76(4.3)\\
 53994.953&     53.6&    $-$4.0$\pm$  4.9&    $-$0.8&     $<0.1$&    6.84(5.8)\\
 53995.961&     58.1&    $-$2.8$\pm$  4.6&    $-$0.6&     $<0.1$&    7.76(6.0)\\
 53996.855&    176.2&     1.6$\pm$  9.1&     0.2&     $<0.1$&   22.15(5.7)\\
 53997.883&    132.7&     5.2$\pm$  7.6&     0.7&     $<0.1$&   22.95(7.8)\\
 54000.852&    165.2&    11.4$\pm$  9.7&     1.2&     $<0.1$&   35.41(9.7)\\
 54002.875&    154.4&    36.8$\pm$ 10.4&     4.0&     3.2&   \nodata       \\ 
 54003.859&    166.9&    $-$7.0$\pm$  9.1&    $-$0.8&     $<0.1$&   13.35(3.6)\\
 54004.891&    123.3&    $-$6.0$\pm$  7.9&    $-$0.7&     $<0.1$&   11.33(4.1)\\
 54005.914&     87.9&    $-$2.2$\pm$  6.3&    $-$0.3&     $<0.1$&   11.88(6.1)\\
 54006.938&     28.0&     5.4$\pm$  4.1&     1.4&     $<0.1$&   15.26(24.6)\\
 54020.891&     65.5&    $-$8.6$\pm$  5.9&    $-$1.4&     $<0.1$&    4.27(2.9)\\
 54021.887&     68.6&    $-$6.2$\pm$  5.7&    $-$1.0&     $<0.1$&    6.30(4.1)\\
 54022.887&     58.1&     1.6$\pm$  5.9&     0.3&     $<0.1$&   14.55(11.3)\\
 54028.863&     68.6&     3.4$\pm$  5.9&     0.6&     $<0.1$&   18.28(12.0)\\
 54029.895&     33.5&     3.4$\pm$  5.1&     0.7&     $<0.1$&   15.93(21.5)\\
 54030.863&     19.6&    $-$1.8$\pm$  3.0&    $-$0.6&     $<0.1$&    5.41(12.5)\\
 54048.824&     47.2&     1.6$\pm$  5.7&     0.3&     $<0.1$&   14.99(14.3)\\
 54049.824&     47.9&    $-$6.0$\pm$  5.4&    $-$1.1&     $<0.1$&    6.09(5.7)\\
 54056.820&     27.1&    $-$5.2$\pm$  3.8&    $-$1.3&     $<0.1$&    3.55(5.9)\\
 54057.820&     21.5&     1.2$\pm$  2.6&     0.5&     $<0.1$&    7.96(16.7)
\enddata
\tablenotetext{a}{From left to right: Modified Julian Date of the
beginning of the observation, total observation EOT, number of excess
events, statistical significance of the excess, equivalent
(\emph{post-trial}) significance for 26 independent samples and signal
upper limit for the different observation nights. A cut SIZE$>$200 photo-electrons
($E_\gamma > 150$ GeV) has been applied. Upper limits
\citep{Rolke2005} are 95$\%$ confidence level (CL) and are quoted in
number of events and in units of the $\gamma$-ray flux measured for
the Crab nebula, assuming the Crab nebula spectral slope ($\alpha=-2.6$).}
\end{deluxetable}

\section{Observations and results}

The BHXB Cygnus X-1 was observed with MAGIC for a total of 46.2 hours
between June and November 2006. MAGIC is an Imaging Atmospheric
Cherenkov Telescope (IACT) located at La Palma (Canary Islands,
Spain), at 28.8$^\circ$N, 17.8$^\circ$W, 2200 m.a.s.l.  The
telescope's sensitivity is $\sim 2\%$ of the Crab nebula flux in 50
hours of observations. The angular resolution is $\sim 0.1^\circ$, and
the energy resolution above 150 GeV is about 20$\%$. MAGIC can provide
$\gamma$-ray source localization in the sky with a precision of $\sim
2^\prime$ and is able to observe under moderate moonlight or twilight
conditions \citep{Albert2007}. At La Palma, Cygnus X-1 culminates at a
zenith angle of $5^\circ$ and the observations were carried out at
zenith angles between 5$^\circ$ and 35$^\circ$. The brightest object
in the Cygnus X-1 field of view is the 3.89 magnitude, K0 spectral-type star
$\eta$~Cygni, located $26^\prime$ away from Cygnus X-1. The
observations were carried out in the false-source track (wobble) mode
\citep{Fomin1994}, with two directions at $24^\prime$ distance and
opposite sides of the source direction. This technique allows for a
reliable estimation of the background with no need of extra
observation time. One of the tracked directions corresponds roughly to
that of $\eta$~Cygni, which reduces the effect of the star in the data
analysis.

\begin{figure}
\epsscale{0.95}
\plotone{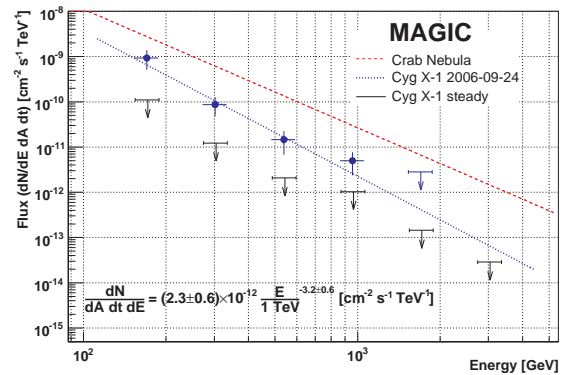}
\figcaption[spectrum_final.eps]{Differential energy spectrum from
Cygnus X-1 corresponding to 78.9 minutes EOT between MJD
54002.928 and 54002.987 (2006-09-24). Also shown are the Crab nebula
spectrum and the best fit of a power-law to the data and the 95$\%$
confidence level upper limits to the steady $\gamma$-ray flux
\citep{Rolke2005}.\label{fig:spectrum}}
\end{figure}

Data corresponding to 46.2 hours from 26 nights of observation were
analyzed using the standard MAGIC calibration and analysis software
\citep{Albert2006b,Gaug2005}. Data runs with anomalous event rates
(6.2 hours) were discarded for further analysis, leading to a total of
40.0 hours of useful data (see Table~\ref{table:log} for
details). Hillas variables \citep{Hillas1985} were combined into an
adimensional $\gamma$/hadron discriminator (\emph{hadronness}) and an
energy estimator by means of the Random Forest classification
algorithm, which takes into account the correlation between the
different Hillas variables
\citep{Breiman2001,Bock2005}. The incoming direction of the primary
$\gamma$-ray events was estimated using the DISP method, suited for
observations with a single IACT \citep{Fomin1994,Domingo2005}. These
algorithms were trained with a sample of Monte Carlo (MC) simulated
$\gamma$-ray events \citep{Majumdar2005} and optimized on 3.7 hours of
observations of the Crab nebula performed during the same epoch at
similar zenith angles (12-32$^\circ$), yielding the signal selection
cuts hadronness$<$0.1 and $\theta<0.1^\circ$ (where $\theta$ is the
angular distance to the source position). The residual background was
evaluated from 5 circular control regions, located symmetrically to
the source position with respect to the camera center. For daily
searches we increase the sample for background estimation by adding
control regions corresponding to close days, obtaining on average 22
times higher statistics than in the on-source region.

\begin{figure}
\epsscale{0.95}
\plotone{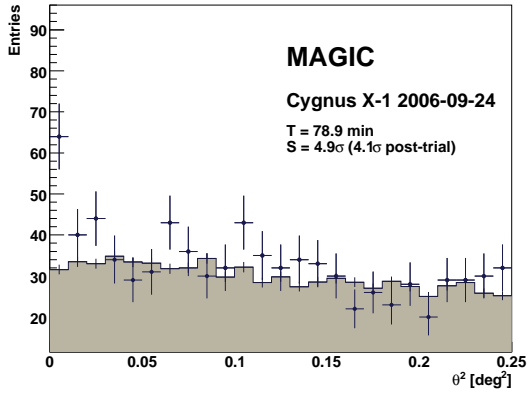}
\figcaption[theta2_final.eps]{Distribution of $\theta^2$ values for
the source (dots) and background (histogram) for an energy threshold
of 150 GeV.\label{fig:theta2}}
\end{figure}

\begin{figure}
\epsscale{0.88}
\plotone{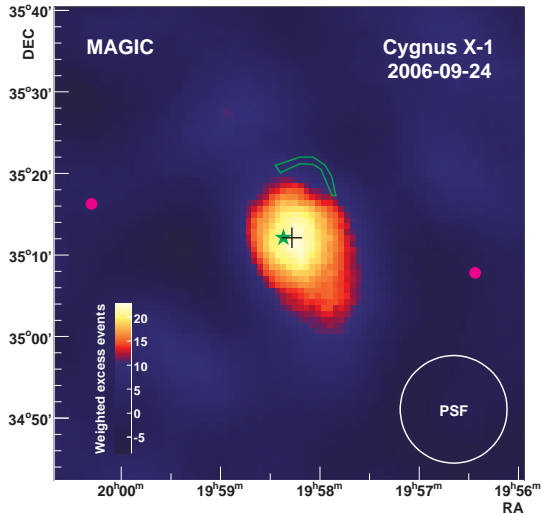}
\figcaption[skymap_final.eps]{Gaussian-smoothed ($\sigma = 4'$) map of
$\gamma$-ray excess events (background subtracted) above 150 GeV
around Cygnus X-1 corresponding to 78.9 minutes EOT between MJD
54002.928 and 54002.987 (2006-09-24). The black cross shows the
best-fit position of the $\gamma$-ray source. The position of the
X-ray source and radio emitting ring-like are marked by the green star
and contour, respectively. The purple dots mark the directions tracked
during the observations. Note that the bin contents are correlated due
to the smoothing.\label{fig:skymap}}
\end{figure}

A search for steady $\gamma$-ray signals was performed for the entire
recorded data sample, yielding no significant excess. This allows us
to establish the first upper limits to the VHE $\gamma$-ray steady
flux of Cygnus X-1 in the range between 150 GeV and 3 TeV (see
Figure~\ref{fig:spectrum}), of the order of 1--5$\%$ of the Crab
nebula flux. Given the time scale of the variability of Cygnus X-1 at
other energy bands, $\gamma$-ray signals are searched for also on a
daily basis. The results are shown in Table~\ref{table:log}.  We
obtain results compatible with background fluctuations at 99$\%$ CL
for all the searched samples except for MJD=54002.875 (2006-09-24). We
derive upper limits to the integral flux above 150 GeV between 2 and
25$\%$ of the Crab nebula flux (depending basically on the observation
time) for all samples compatible with background fluctuations. The
data from 2006-09-24 were further subdivided into two halves to search
for fast varying signals, obtaining 0.5$\sigma$ and 4.9$\sigma$
effects for the first (75.5 minutes EOT starting at MJD 54002.875) and
second (78.9 minutes EOT starting at MJD 54002.928) samples,
respectively. The post-trial probability is conservatively estimated
by assuming 52 trials (2 per observation night) and corresponds to a
significance of 4.1$\sigma$.  The sample corresponding to MJD
54002.928 was further subdivided into halves, obtaining 3.2$\sigma$
and 3.5$\sigma$ excesses in each. At this point we stopped the data
split process.

The distribution of $\theta^2$ for signal and background events
corresponding to the 78.9 minutes EOT sample starting at MJD 54002.928
is shown in Figure~\ref{fig:theta2}. The excess is consistent with a
point like source located at the position of Cygnus X-1. The map of
excess events around the source is shown in Figure~\ref{fig:skymap}. A
Gaussian fit yields the location: $\alpha = 19^\textrm{h}
58^\textrm{m} 17^\textrm{s}$, $\delta = 35^\circ 12' 8''$ with
statistical and systematic uncertainties $1.5^\prime$ and $2'$,
respectively, compatible within errors with the position of Cygnus X-1
and excluding the jet-powered radio nebula at a distance of $\sim
8'$. The energy spectrum is shown in Figure~\ref{fig:spectrum}. It is
well fitted ($\chi^2/n.d.f=0.5$) by the following power law:
$dN/(dA~dt~dE) = (2.3\pm0.6)\times 10^{-12} (E/1~\textrm{TeV})^{-3.2\pm 0.6}
\textrm{cm}^{-2} \textrm{s}^{-1} \textrm{TeV}^{-1}$ where the quoted
errors are statistical only. We estimate the systematic uncertainty to
be $35\%$ on the overall flux normalization and 0.2 in the
determination of the spectral index.

\section{Discussion}

\begin{figure}
\epsscale{0.7}
\plotone{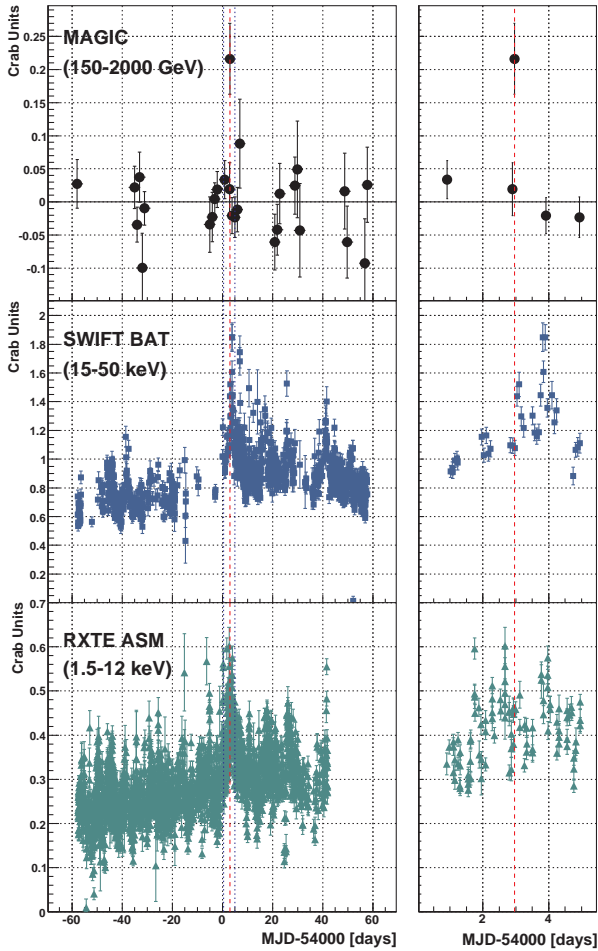}
\figcaption[lighcurve_final.eps]{From top to bottom: MAGIC, {\it Swift}/BAT
(from http://swift.gsfc.nasa.gov/docs/swift/results/transients/) and
{\it RXTE}/ASM (from http://heasarc.gsfc.nasa.gov/xte\_weather/)
measured fluxes from Cygnus X-1 as a function of the time. The left
panels show the whole time spanned by MAGIC observations. The
vertical, dotted blue lines delimit the range zoomed in the right
panels. The vertical red line marks the time of the MAGIC
signal.\label{fig:lightcurve}}
\end{figure}

\setcounter{footnote}{0}

The excess from the direction of Cygnus X-1 occurred simultaneously
with a hard X-ray flare detected by {\it INTEGRAL} ($\sim$1.5~Crab
between 20--40~keV and $\sim$1.8~Crab between 40--80~keV)
\citep{Turler2006}, {\it Swift}/BAT ($\sim$1.8 Crab between 15 and 50
keV) and {\it RXTE}/ASM ($\sim$0.6 Crab between 1.5 and 12
keV). Figure~\ref{fig:lightcurve} shows the correlation between MAGIC,
{\it Swift}/BAT and {\it RXTE}/ASM light-curves. The TeV excess was
observed at the rising edge of the first hard X-ray peak, 1--2 hours
before its maximum, while there is no clear change in soft
X-rays. Additionally, the MAGIC non-detection during the following
night (yielding a $95\%$ CL upper limit corresponding to a flux $\sim
5$ times lower than the one observed in the second half of 2006-09-24)
occurred during the decay of the second hard X-ray peak. This
phenomenology leads us to think that, during the 2006-09-24 night,
soft and hard X-rays are produced in different regions. Furthermore,
hard X-rays and VHE $\gamma$-rays could be produced at regions linked
by the collimated jet, e.g.\ the X-rays at the jet base and
$\gamma$-rays at an interaction region between the jet and the stellar
wind. These processes would have different physical timescales, thus
producing a shift in time between the TeV and X-ray peaks. Note that
the distance from the compact object to the TeV production region is
constrained below 2$^\prime$ by MAGIC observations and therefore it is
unrelated with the nearby radio emitting ring-like structure
\citep{Gallo2005}. A jet scenario is, however, not devoid of
constraints either. The observed TeV excess took place at phase 0.91,
being 1 the moment when the BH is behind the massive star. At this
phase, MAGIC observations are available only for the night 2006-09-24,
which precludes any possible analysis of a putative periodicity
feature of the TeV emission. If the TeV emission were produced in the
jet well within the binary system, the photon-photon absorption in the
stellar photon field would be dramatic, yielding a TeV detection very
unlikely. For instance, \citet{Bednarek2007} computed the opacity to
pair production for different injection distances from the center of
the massive star and angles of propagation, finding that photons
propagating through the intense stellar field towards the observer
would find in their way opacities of about 10 at 1 TeV. Admittedly,
inclination of the orbit and angle of propagation to the observer can
change these numbers, but not the fact that MAGIC observes the excess
at the position where the expected opacity is highest. Therefore, even
without an explanation for a TeV flare, we must consider that the
emission could have been originated far from the compact
object. Interactions of the jet with the stellar wind may lead to such
a situation.

In summary, for the first time we have found experimental evidence of
VHE emission produced by a Galactic stellar-mass BH. It is also the
first evidence of VHE gamma-rays produced at an accreting binary
system. Our results show that a possible steady VHE flux is below the
present IACT's sensitivity and tight upper limits have been
derived. On the other hand, we have found evidence for an intense
flaring episode during the inferior conjunction of the optical star,
of time scale shorter than 1 day and rising time of about 1 hour,
correlated with a hard X-ray flare observed by {\it Swift} and {\it
INTEGRAL}. These results point to the existence of a whole new
phenomenology in the young field of VHE astrophysics of binary systems
to be explored by present and future IACT's.


We thank the IAC for the excellent working conditions at the
Observatory del Roque los Muchachos in La Palma.
The support of
the German BMBF and MPG, the Italian INFN and the Spanish CICYT is
gratefully acknowledged. This work was also supported by ETH
Research Grant TH~34/04~3 and the Polish MNiI Grant 1P03D01028.


\end{document}